\documentclass{kluwer}    
\usepackage{graphicx}

\begin{document}
\begin{article}
\begin{opening}         
\title{Cluster--Galaxy Interactions in Coma}
\author{Michael D. \surname{Gregg}, Bradford P. \surname{Holden}, and}  
\runningauthor{Michael D. Gregg}
\runningtitle{Coma Galaxy Interactions}
\institute{Univ.\ of California, Davis, and Inst. for Geophysics and
            Planetary Physics, Lawrence Livermore National Laboratory}
\author{Michael J. \surname{West}}
\institute{Univ.\ of Hawaii, Hilo}

\begin{abstract}

The galaxies NGC 4911 and 4921 in the Coma cluster provide spectacular
examples of luminous, grand design spirals passing through the hot
intracluster medium of a rich cluster of galaxies.  Chandra ACIS-I
imaging of their environment, coupled with CFHT, HST/WFPC2, and VLA
radio imaging, reveal the interactions between the hot X-ray gas and
the ISM of the spirals, highlighting the linkage between galaxy and
cluster evolution generated by such interactions.                     
\end{abstract}

\end{opening}           

\section{Introduction}  

Coma, like most large clusters of galaxies, is still being built
through infall of subclusters and smaller groups.  Evolution of the
infalling galaxies is altered by the cluster environment where hot
intracluster gas can ablate or strip the interstellar medium of
galaxies, radically altering their star formation histories.  Systems
can be ``harassed'' (Moore et al.\ 1996) or, in the extreme, even
completely disrupted by tidal forces, spilling their contents
throughout intracluster space (Gregg \& West 1998).  Infalling gas
stripped from galaxies augments and heats the intracluster medium
(ICM), thereby accelerating galaxy evolution through ram-pressure
stripping.  Over time, the material removed from galaxies is recycled
into an ever-growing intracluster population of stars, gas, and dwarf
galaxies (e.g., West et al.\ 1995; C\^ot\'e, Marzke \& West 1998;
C\^ot\'e et al.\ 2000; Thompson \& Gregory 1993; Lopez-Cruz et al.\
1997).

There are two giant spiral galaxies in the core of Coma, NGC4911 and
NGC4921, which are undergoing intense and stressful interactions with
their environment.  Gregg \& West (1998) documented several extended
low surface brightness (LSB) objects in the Coma cluster, one of which
is trailing NGC4911 into the core of Coma (Figure~1).  Such LSB
features are most likely transient, produced by galaxy-galaxy
interactions or by stripping of galaxies by the global cluster tidal
field and ICM.

\section {Optical Observations}

We have obtained HST/WFPC2 V (F606W) and I (F814W) images of NGC~4911
and also of its LSB tidal debris wake.  The left panel in Figure~1
shows the original discovery image of the LSB material following in
the wake of NGC~4911; at this contrast, the galaxy and its companion
S0 are burned into one.  The center panel shows the WFPC2 V image of
NGC~4911 at intermediate contrast to reveal the companion; the WF
chips are filled with extended low surface brightness material from
NGC4911.  The right panel shows just the PC chip to reveal the nuclear
region; there appears to be a disembodied spiral arc of enhanced
surface brightness on the leading side of the disk toward the cluster
core, which is $0.29^{\circ}$ away.  This arc may be an example of 
ram-pressure-induced star formation, perhaps a bow shock, as NGC~4911
plows through the hot X-ray emitting ICM in Coma.


\begin{figure}[H]
\includegraphics[width=5in]{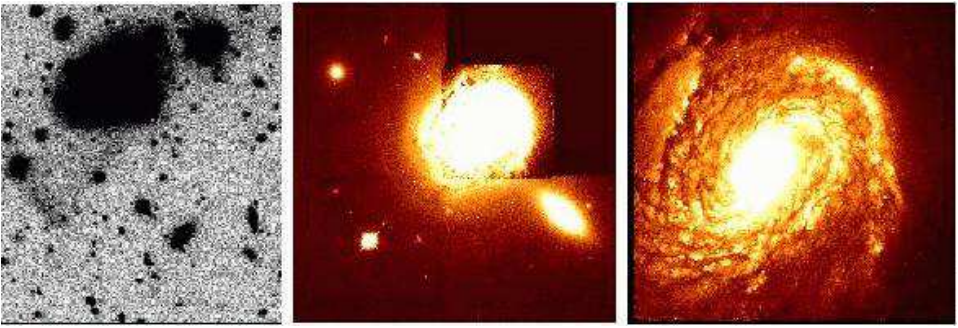}
\caption { {\bf Left:} A portion of the KPNO Schmidt discovery image
of the LSB tidal debris following NGC~4911.  {\bf Center:} WFPC2 F606W
(V-band) image of NGC4911.  The S0 galaxy (NGC4911A) to the lower right
is perhaps responsible for generating the tidal debris, but itself
shows no morphological disturbances.  {\bf Right:} F606W (V-band) PC
field shown at different stretch to reveal more detail.  The core of
Coma is to the upper right; on this side of the galaxy is a bright,
disjoint spiral arm, perhaps the result of ram pressure-induced star
formation.}
\end{figure}


The second WFPC2 field was obtained with most of the trailing LSB
material on WF3, clearly seen in {Figure~2}.  Using {\sc sextractor},
we have cataloged all objects in the LSB image; a 3$\sigma$ excess of
objects is found in the WF3 chip relative to the other two WFPC2
fields.  These objects have roughly globular cluster luminosities;
if these are really in Coma, we are witnessing the present-epoch
creation of intergalactic star clusters.  We will be obtaining
redshifts with Keck for the brightest objects in this field; a
spectrum obtained last year shows that one of the brightest has
emission lines and a velocity placing it in the cluster.

\begin{figure}[t!]
\caption {WFPC2/WF3 image of LSB material near NGC4911.  Circles
indicate Chandra X-ray point sources.}
\includegraphics[width=3.5in]{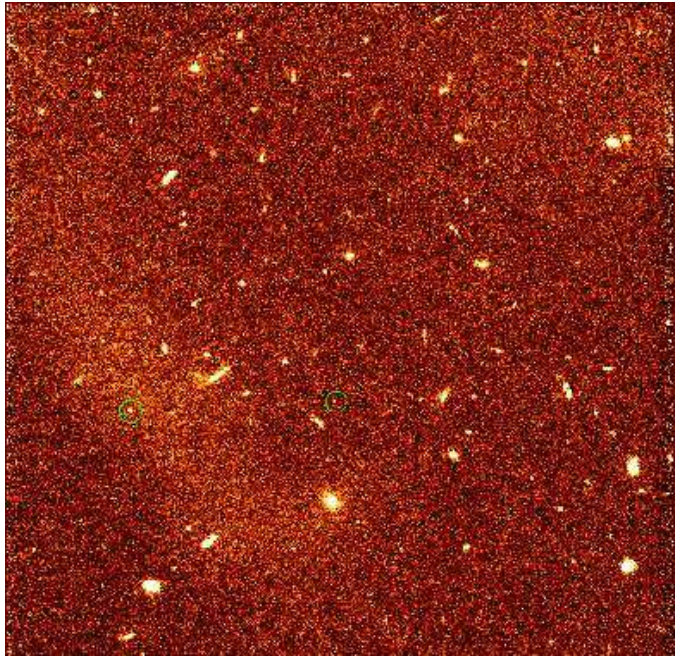}
\end{figure}

\section {X-ray Observations}

We obtained a 65ks Chandra image, centered near NGC4911, a known X-ray
sources.  Our Chandra X-ray spectrum of NGC4911 shows that it houses
an active galactic nucleus, perhaps spawned by the same interactions
which have produced the LSB material.

Of the many point sources in our Chandra field, two have optical
counterparts in the WF3 field of the LSB material ( {circles} in
Figure~2).  The brighter of these has a soft X-ray spectrum,
consistent with being an X-ray binary in Coma with a luminosity of
$\sim 7\times10^{38}$ ergs/s.  The other two WFPC2 chips have no X-ray
point sources with optical counterparts.

\section {Radio Observations}

VLA 21cm observations also show evidence for interactions between NGC~4911
and the intracluster medium (Figure~3).  The VLA observations also
captured the continuum of NGC4921, a giant spiral to the east of
NGC4911.  Both galaxies appear
to be losing their ISM, as indicated by the distorted, asymmetric, and
flowing contours seen in the radio (Figure~3).  More
extensive studies by Bravo-Alfaro et al.\ (2000) show similar behavior
for these galaxies and other, smaller spirals in Coma.

\begin{figure}[t!]
\includegraphics[width=4.5in]{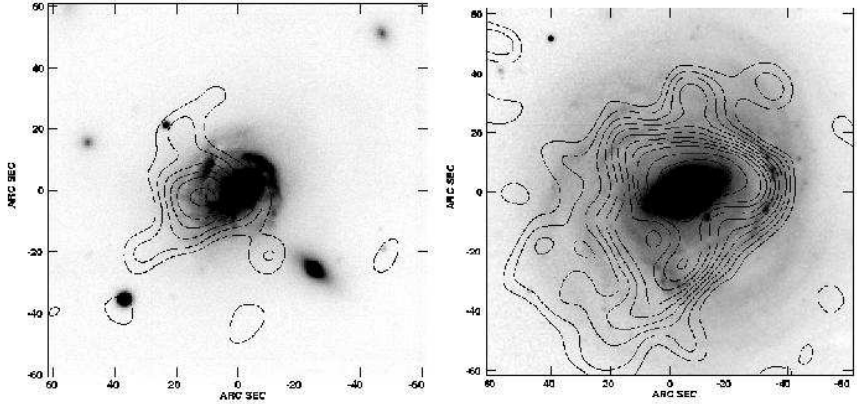}
\caption{{\bf LEFT:} VLA C-array map of HI in NGC~4911 for a
representative velocity channel, superimposed on a CFHT R-band image.
The gas is off-center and exhibits a confused velocity structure,
evidence of being stripped from the galaxy by the hot ICM.  {\bf
RIGHT:} A VLA C-array 20cm continuum map of NGC~4921, superimposed on
a CFHT image.  The gas is being compressed on the side towards the
cluster core and extends much farther out into the disk on the opposite side,
evidence of being stripped from the galaxy.  Cluster core is to the
upper right in both panels.}
\end{figure}

\section {Summary}
\begin{itemize}
\item {Coma, dominated by early type galaxies, is still accreting
spiral galaxies at the present epoch.}
\vspace {-0.1in}
\item {The spirals are being drastically transformed by their
interaction with the cluster.  Vigorous star formation is induced,
using up the gas, and the ICM also
removes gas by heating and stripping the galaxies' ISM.}
\vspace {-0.1in}
\item {The outer stellar portions of NGC4911 are being removed by 
the global cluster tidal field or its S0 companion, further
altering its evolution.}
\vspace {-0.1in}
\item {The stripped material will
eventually disburse throughout the core of Coma, augmenting the
general intergalactic populations of stars, star clusters, gas, and
perhaps even X-ray binaries.  Through this process, the infalling
galaxies in turn help drive the evolution of the whole cluster.}
\end{itemize}

\acknowledgements

The authors acknowledge generous support from the National Science
Foundation (AST~9970884), and NASA through STScI and the Chandra X-ray
Center.  Part of the work reported here was done at IGPP, under the
auspices of the U.S. Department of Energy by Lawrence Livermore
National Laboratory under contract No.~W-7405-Eng-48.


\end{article}
\end{document}